\documentclass[conference,a4paper]{IEEEtran}
\usepackage{cite}
\usepackage{amsmath,amssymb,amsfonts}
\usepackage{algorithmic}
\usepackage{graphicx}
\usepackage{textcomp}
\usepackage{subcaption}
\usepackage{soul}
\usepackage{graphicx}
\usepackage{xcolor}

\def\BibTeX{{\rm B\kern-.05em{\sc i\kern-.025em b}\kern-.08em
    T\kern-.1667em\lower.7ex\hbox{E}\kern-.125emX}}
\begin{document}

\title{Probability of Line of Sight Evaluation in Urban Environments using 3D Simulator
}

\author{Abdul Saboor$^{1,*}$,
        Evgenii Vinogradov\textsuperscript{1, 2},
        Zhuangzhuang Cui\textsuperscript{1},
        Sofie Pollin\textsuperscript{1}
\\\textsuperscript{1}Department
of Electrical Engineering, KU Leuven, Belgium
\\\textsuperscript{2}Autonomous Robotics Research Center, Technology Innovation Institute, UAE
\\Email$^{*}$: abdul.saboor@kuleuven.be
}

\maketitle

\begin{abstract}
The integration of Non-Terrestrial Networks (NTNs) into 6G networks is one of the most promising ways to achieve significant improvements in capacity, reliability, and global coverage. The design of NTN heavily relies on using channel models. In this paper, we propose two easy-to-use simulators for estimating the Line-of-Sight (LoS) probability $\mathbf{P_{LoS}}$ in a 3D urban environment. The first simulator is a 3D city simulator that employs simplified Ray Tracing (RT), while the second one is a lightweight geometry-based simulator that generates only the relevant buildings between users and Unmanned Aerial Vehicles (UAVs). Using these simulators, we assess the accuracy of existing models for $\mathbf{P_{LoS}}$ estimation and examine $\mathbf{P_{LoS}}$ for different UAV heights, user-UAV distances, and azimuth/elevation angles. We conclude that 1) existing models overestimate $\mathbf{P_{LoS}}$, resulting in overoptimistic NTN performance predictions, 2) nodes location (including azimuth and elevation angles) is an important factor influencing $\mathbf{P_{LoS}}$, however, this influence is not captured by the existing models. 

\end{abstract}

\begin{IEEEkeywords}
Unmanned Aerial Vehicles (UAV), Probability of Line Of Sight ($\mathbf{P_{LoS}}$), Aerial Base Station (ABS).
\end{IEEEkeywords}
\section{Introduction}
The next generation of wireless networks, known as 6G, is expected to bring significant advancements for capacity, reliability, and global coverage \cite{9178307}. However, to achieve these improvements, new technologies and architectures will need to be developed. One promising 6G enabler is the integration of Non-Terrestrial Networks (NTNs) into 6G. NTNs, which include satellite, high- and low-altitude aerial networks, can provide high-speed connectivity to remote areas and support new use cases such as autonomous vehicles, remote surgery, and smart agriculture \cite{shrestha20216g}.

Multiple research contributions have been dedicated to describing the benefits of NTN-enabled 6G networks \cite{9178307,9992172,8918497}. Indeed, Unmanned Aerial Vehicles (UAVs, known as drones) can carry an Aerial Base Station (ABS) that can provide better and on-demand coverage than Terrestrial Base Stations by tuning its mobility in the three-Dimensional (3D) space \cite{id6,id7}. Moreover, High-altitude Platform Stations (HAPS) can ensure seamless connection for very long distances \cite{9380673} enabling next-generation intelligent inter-continental transportation \cite{9815183}. Additionally, the airspace is increasingly inhabited by various cellular-connected agents \cite{9768113} such as UAVs performing their missions, e.g., package delivery \cite{achiel} or even heavy cargo \cite{id6}, or Reflective Surfaces used to boost NTN performance \cite{9772693}. 

All these exciting use cases are motivated by the promise of a stable radio channel offering high-capacity communication due to higher Line-of-Sight (LoS) availability when compared with conventional terrestrial communications.
In this context, channel models become essential for designing and analyzing the performance of the corresponding communication systems \cite{id8}. One critical part of U2G channel models is LoS availability, which significantly impacts communication performance. In contrast to the Non-LoS (NLoS) links experiencing shadowing due to physical objects (buildings, trees), the radio waves in the LoS link propagate better thanks to the clear scattering environment between the transmitter and receiver. Hence, the channel reliability and transmission rates are directly associated with the probability of LoS $P_{LoS}$. 

\begin{figure}[!t]
\centerline{\includegraphics[width=.8\linewidth]{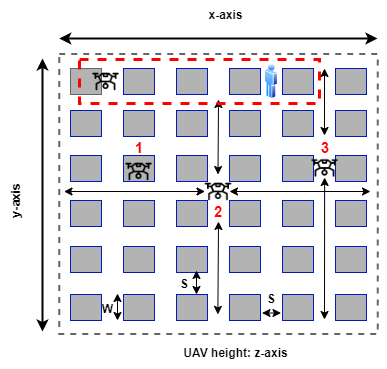}}
\caption{A top view of Manhattan urban environment. }
\label{fig1}
\end{figure}

In existing works, $P_{LoS}$ is evaluated by measurement-based empirical \cite{3GPP} and stochastic/deterministic geometry-based modeling \cite{ITU, achiel}. When measured data (LoS or geographical information about the considered area) is not available, a Manhattan grid structure (see Fig.~\ref{fig1}) is frequently used to model the city layout. Based on this reference geometry, \cite{id13,ICC} propose empirical $P_{LoS}$ models using the fitting parameters to match the simulator and model results. Unfortunately, these models have several disadvantages: 1) the empirically estimated fitting parameters are valid only for a limited set of scenarios and 2) the models are, in fact, two-dimensional (x-axis and z-axis) where the elevation angle between the UAV and user is considered in a straight line, as illustrated by the red box in Fig. \ref{fig1}. In reality, the user or UAV can be anywhere in a 3D city, and the location of the UAV directly impacts $P_{LoS}$ \cite{9419751}. For example, the UAV on a crossroad would have better $P_{LoS}$ compared to a UAV located in the street or on top of the building, as illustrated in Fig. \ref{fig1}. \cite{9419751} is free of these disadvantages, however, it is based on an alternative city geometry which makes it difficult to compare with the standard model \cite{ITU} and its derivatives \cite{id13,ICC}. 

To overcome the aforementioned disadvantages of the available models, we propose easy-to-use, lightweight, and real-time city simulators. The first simulator can create a 3D city using any built-up parameters having a different number of buildings, heights, and area with random or predefined UAVs and users' locations for estimating $P_{LoS}$. In contrast, the second simulator is a lightweight geometry-based simulator that randomly locates users and UAVs in the city and generates only buildings relevant to the considered line between users and UAVs. Next, we use the simulators to assess the accuracy of the models in \cite{id13,ICC}. The overall contributions to this paper are listed as follows:  

\begin{itemize}
\item We develop a 3D city simulator using simplified Ray Tracing (RT) for the $P_{LoS}$ evaluation in any environment that can be 
reconfigured based on three generic built-up parameters defined by ITU in \cite{ITU}.
\item We propose a lightweight geometry-based method to validate the results of the 3D city simulator. In contrast to other 3D simulators, the geometry-based simulator is computationally less expensive and it is scalable to relatively large cities. Therefore, it is beneficial for estimating $P_{LoS}$ at low elevation angles.
\item We examine the $P_{LoS}$ against UAV height, distance from users, and azimuth/elevation angles between the ground user and UAV using both 3D and geometry-based simulators. 
\end{itemize}

The rest of the paper is organized as follows: Section II introduces the city layout and built-up parameters. Section III presents both simulators for LoS evaluation. Section IV illustrates the results, and Section V concludes the paper.

\section{City Layout}
This work is based on the Manhattan grid type urban city model \cite{ITU}. The three built-up parameters include 
\begin{itemize}
    \item \textbf{$\alpha$}: The ratio of land area covered by buildings over the built-up/city area,
    \item \textbf{$\beta$}: Average buildings per unit area (buildings/$km^2$),
    \item \textbf{$\gamma$}: Rayleigh parameter that describes the distribution of the building heights in different city environments.
\end{itemize}

\begin{table}[!t]
\caption{Built-up Parameters for Typical Environments.}
\begin{center}
\begin{tabular}{|p{2cm}|p{1.5cm}|p{2cm}|p{1.5cm}|}
\hline 
\textbf{Environment} & \textbf{$\alpha$} & \textbf{$\beta$ (buildings/$km^2$)}  & \textbf{$\gamma$ (m)} \\ \hline
\textbf{Suburban} & 0.1 & 750 & 8 \\ \hline
\textbf{Urban} & 0.3 & 500 & 15 \\ \hline
\textbf{Dense Urban} & 0.5 & 300 & 20 \\ \hline
\textbf{High-rise Urban} & 0.5 & 300 & 50 \\ \hline
\end{tabular}
\label{tab1}
\end{center}
\end{table}

\begin{figure}[!t]
\centerline{\includegraphics[width=.8\linewidth]{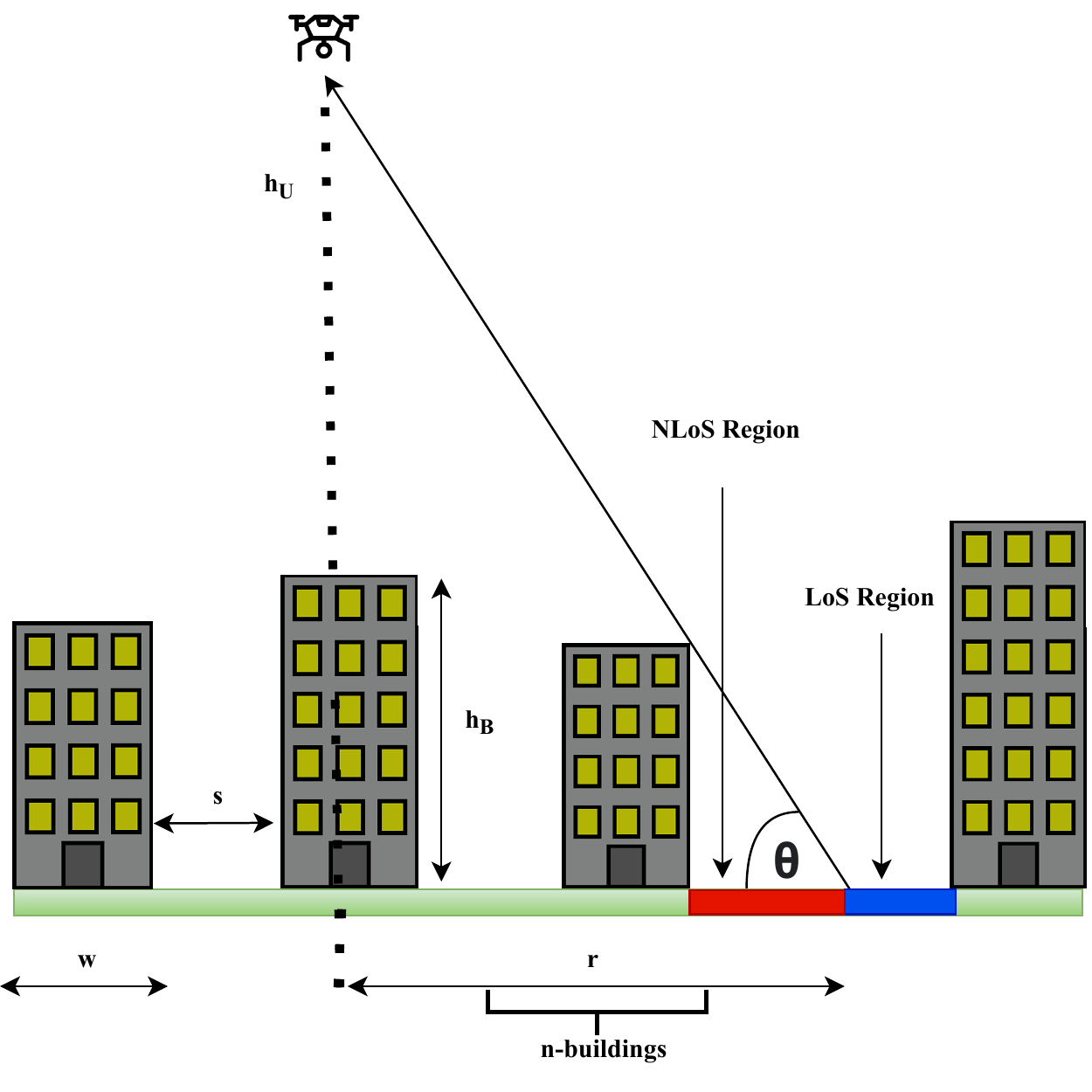}}
\caption{The LoS and NLoS regions for a certain $\theta$.}
\label{fig2}
\end{figure}
The built-up parameters \cite{ITU} for four typical urban city environments are illustrated in Table \ref{tab1}. It only requires $\alpha$, $\beta$ and $\gamma$ to generate different city environments. Also, any generalized city model can be created by using or changing these built-up parameters. Here, $\gamma$ is the scale parameter of Rayleigh distribution that defines the probability distribution of buildings' height in urban environments. The building heights ($h_B$) can be generated based on Rayleigh distribution whose Probability Density Function (PDF) is expressed by 
\begin{equation}
\label{eq1}
    f(h_B) = \frac{h_B}{\gamma^2}e^{-\frac{h_B^2}{2\gamma^2}}.
\end{equation}

In a grid urban, the buildings are assumed to be square with a fixed size ($w$). Furthermore, all the buildings are equally spaced, known as streets ($s$), for an exemplified environment shown in Fig. \ref{fig1}. The streets are structure-free places in the city, including roads, pavements, and gardens. The values of $s$ and $w$ depend on the environment and are calculated by \cite{id17} 
\begin{equation}
\label{eq2}
    w = 1000  \sqrt{\frac{\alpha}{\beta}},
\end{equation}

\begin{equation}
\label{eq3}
    s = \frac{1000}{\sqrt{\beta}} - w.
\end{equation}

\subsubsection{LoS/NLoS definition}
Suppose a UAV is flying above the city at an altitude $h_U$. The communication link is considered LoS if there is no building obstructing the line between the UAV and the user. 
In contrast, the link is NLoS if a building comes between the UAV and the user.

The elevation angle ($\theta$) between the UAV and users is critical for $P_{LoS}$ estimation, as emphasized by SOTA studies \cite{id13, id17, ICC}. It also helps to identify LoS and NLoS regions in the street. Fig. \ref{fig2} illustrates the LoS and NLoS regions for a particular $\theta$. Suppose a UAV is flying at an altitude $h_U$. Then one or multiple buildings will disrupt the LoS link between the UAV and users in the street for a particular elevation angle $\theta$, as illustrated in the figure. We term this the shadow of a building that disrupts the direct LoS link. Hence, there is no LoS connection in the red region of the street in the figure. In contrast, all the users in the non-shadow blue region will have the LoS connection. Naturally, the shadow proportion depends on the elevation angle, street width and building heights.  

Another important point to be considered in this paper is the placement of the UAV. Generally, UAV's location can be classified into three major areas (illustrated in Fig. \ref{fig1}):
\begin{enumerate}
    \item Top of the building,
    \item Crossroads,
    \item Street (between two buildings).
\end{enumerate}

The location directly impacts the $P_{LoS}$, as visualized in Fig.~\ref{fig1}. The UAV above the crossroads will have full LoS coverage for four streets. Similarly, UAV flying above the streets will have complete LoS coverage for the users on that street. Hence, such placements will theoretically have better $P_{LoS}$ and must be evaluated. Therefore, one primary objective of this study is to examine the significance of such UAV locations on $P_{LoS}$. Therefore, we present a 3D simulator thoroughly discussed in the next Section.  

\section{Proposed Simulators}
This section presents our proposed 3D city simulator and the lightweight geometry-based simulator.

\subsection{3D City Simulator}
\begin{figure}[!t]
\centerline{\includegraphics[width=0.8\linewidth]{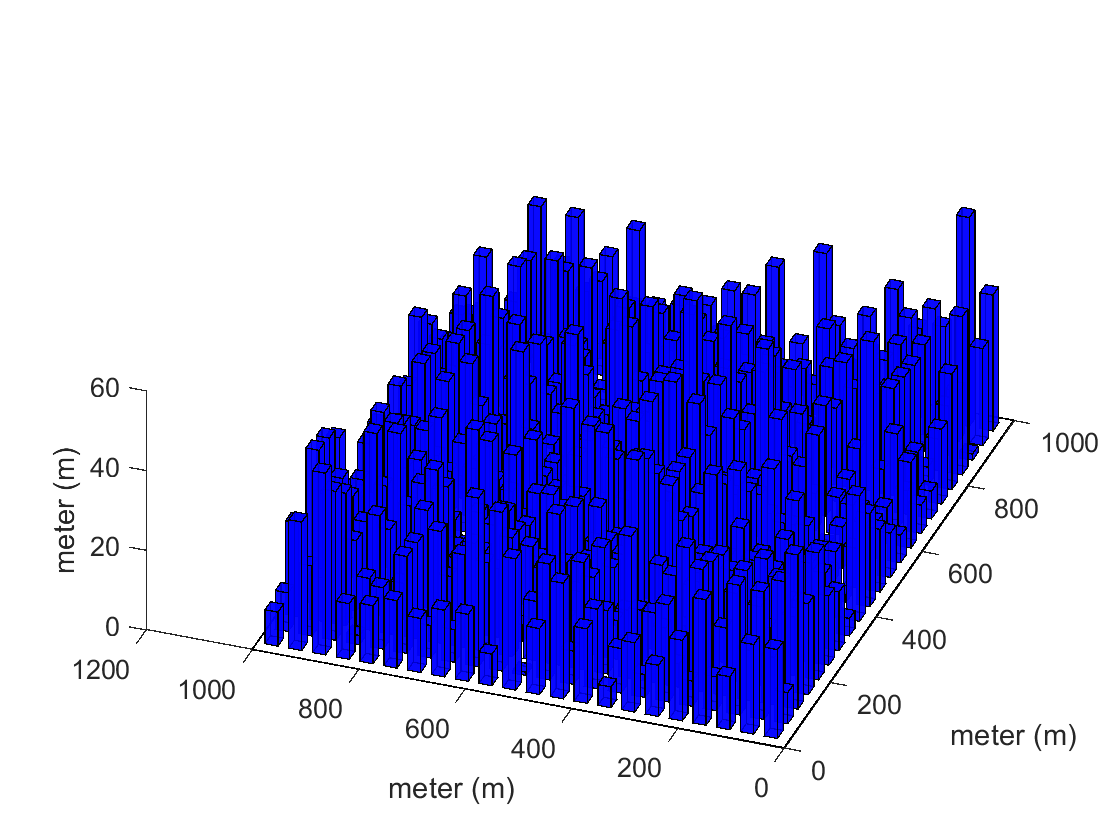}}
\caption{Urban city generated via 3D city simulator. }
\label{fig3}
\end{figure}

The 3D city simulator can generate different cities using built-up parameters, as shown in Fig. \ref{fig3}. The main task of the 3D simulator is to calculate the $P_{LoS}$ for a certain height and $\theta$ between UAV and ground users. For that, we calculate the ray height ($h_{LoS}$) at the $i$-th Obstructing Point (OP) using   
\begin{equation}
\label{ITU}
    h_{LoS(i)} = h_{U} - \frac{r_{OP(i)}*(h_{U}-h_{rx})}{r_{rx}}.
\end{equation}

\begin{figure}[!t]
\centerline{\includegraphics[width=0.85\linewidth]{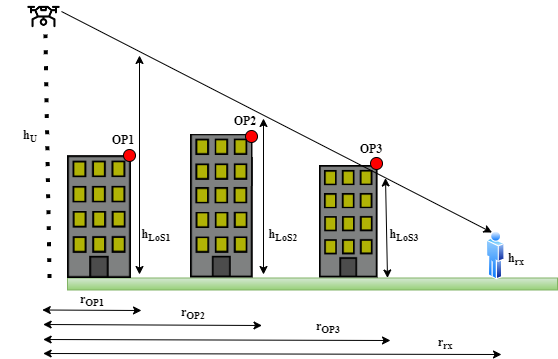}}
\caption{An illustration of the calculation of the ray height at the OP for LoS evaluation in the 3D city simulator.}
\label{fig4}
\end{figure}
Fig. \ref{fig4} illustrates the calculation of ray height at the OP for the LoS evaluation. Suppose a UAV, flying at an altitude of ($h_{U}$), communicates with a user at $r_{rx}$ distance. The height of the receiver is $h_{rx}$. There exist three buildings between the UAV and the user at the distances of $r_{OP1}$, $r_{OP2}$ and $r_{OP3}$, respectively. The heights of the buildings are not identical because all the buildings are Rayleigh distributed. The UAV and user will only be in LoS state if the ray's height is more than all the OPs in the way, three buildings in this case. Instead of evaluating the ray height against each pixel or meter of the building, we are only considering the edge point or OP of the building towards the user. As we consider flat rooftops, if the link is LoS at OP1, it cannot become NLoS due to an obstruction caused by the first building. Therefore, it reduces the overall computational complexity by checking the single OP of each building, rather than checking all the points of the building. 

OP1, OP2, and OP3 are the edge points for each building, as shown in Fig. \ref{fig4}. Therefore, the simulator will start calculating the ray height and compare it with the OP height (building height) at OP1 using \eqref{ITU}. If OP1 height exceeds the ray height, the simulator will consider this ray as NLoS and stop checking for the remaining buildings. In contrast, if ray height is more than OP1, it will examine the next OP2, as shown in Fig \ref{fig4}. Henceforth, the simulator will compare all the edge OPs and ray heights between the UAV and user. The channel will only be in the LoS state if the ray height exceeds all OPs. Otherwise, the state is NLoS, as illustrated in OP3 of the figure. 
\\
\textbf{Advantages}
\begin{itemize}
    \item The proposed 3D simulator graphically constructs an entire city with buildings and streets. It can calculate the $P_{LoS}$ against any $h_U$, $h_{rx}$, $\theta$ and $r_{rx}$ between the UAV and user, in any direction or location of the city.
    \item It can examine the $P_{LoS}$ at different or predefined UAV locations.
    \item It can measure $P_{LoS}$ for different trajectories of users in any city environment based on built-up parameters.
\end{itemize}
\textbf{Disadvantages} 
\begin{itemize}
\item The 3D simulator generates a new city for every simulation. The Big O complexity of the built-up area is $\mathcal{O}(\rho w^2 \beta)$, where $\rho$ is the total land parameter represented by $km^2$. It means the simulator will create $\rho w^2 \beta$ times built-up structure for every simulation. Generally, $n = \bigg\lfloor \frac{h_M}{\tan(\theta)(S+W)} \bigg\rfloor$ buildings obstruct LoS, where $n \ll \rho w^2 \beta$. Therefore, creating such unnecessary buildings make the simulator computationally expensive.    
\end{itemize}

\subsection{Geometry-based Simulator}

\begin{figure}[!t]
\centerline{\includegraphics[width=.85\linewidth]{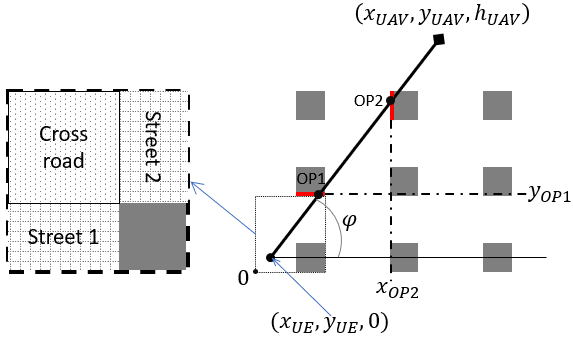}}
\caption{Overview of the geometry-based simulator. }
\label{fig5}
\end{figure}

The geometry-based simulator does not generate the whole city. First, it defines the users and UAV positions and draws the line between these points. Next, it checks if this line crosses buildings in the XY plane. Finally, the building heights are generated, and we check whether the buildings are lower than $h_{LoS}$, analogously to the 3D city simulator.
The procedures of this simulator involve the following steps:
\begin{enumerate}
    \item The user is uniformly placed at the streets of the crossroad, see Fig.~\ref{fig5}. 
    \item For a given $\theta$, the azimuth angle $\varphi$ is uniformly distributed in $[0,90^{\circ}]$ (we exploit the symmetry of the city layout). The UAV height is also generated from a Uniform distribution $\mathcal{U}_{[0,500 m]}$ to account for expected ABS deployment altitudes. Based on the user location $(x_{UE},y_{UE},0)$, UAV deployment height $h_{UAV}$, and angles $\theta$ and $\varphi$, the calculate location of the UAV $(x_{UAV},y_{UAV},h_{UAV})$.
    \item Since we consider only the first quadrant of the XY plane (see Step 2 for $\varphi$), the line of sight can "hit" only two sides of buildings, indicated by the red color in Fig.~\ref{fig5}. Consequently, to find OPs, we check intersections of the line user-UAV and $x_i=(i-1)\cdot(s+w)+s$ and $y_j=j\cdot(s+w)$, where $i,j$ denote indexes of buildings (for example, in the figure, OP2 has $i=j=2$ and OP1 has $j=1$.) 
    \item For each OP, $h_{LoS(i,j)}$ is calculated and compared with a random variable generated from Rayleigh distribution defined by parameter $\gamma$. If any of the buildings $i,j$ is higher than $h_{LoS(i,j)}$, the link is marked as NLoS.  
\end{enumerate}

\textbf{Advantages}
\begin{itemize}
  \item The geometry-based simulator generates only a relevant city portion for evaluation. Hence, removing unnecessary buildings makes the geometry-based simulator lightweight and scalable. The Big O complexity of the built-up area in this simulator is $\mathcal{O}(nm)$, where $m$ is the number of users in a city. In contrast, the complexity of the 3D simulator is independent of $m$.   
    \item It works efficiently even if the distance between users and UAV is long. 
\end{itemize}
\textbf{Disadvantages} 
\begin{itemize}
    \item It cannot define specific UAV locations in a city for $P_{LoS}$ examination. Therefore, it cannot examine the impact of different UAV locations on $P_{LoS}$. 
\end{itemize}
\begin{figure*}[!t]
  \centering
    \begin{subfigure}[b]{0.3\linewidth}
    \includegraphics[width=\linewidth]{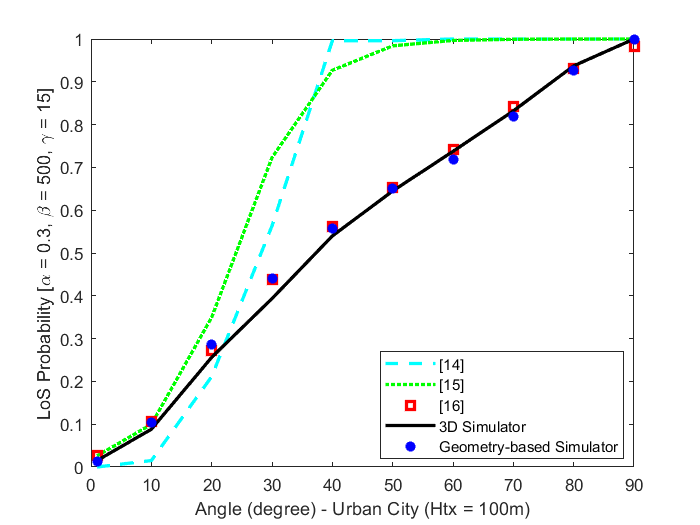}
    \caption{$P_{LoS}$ vs. $\theta$ (urban).}
    \label{fig61}
  \end{subfigure}
  \centering
  \begin{subfigure}[b]{0.3\linewidth}
    \includegraphics[width=\linewidth]{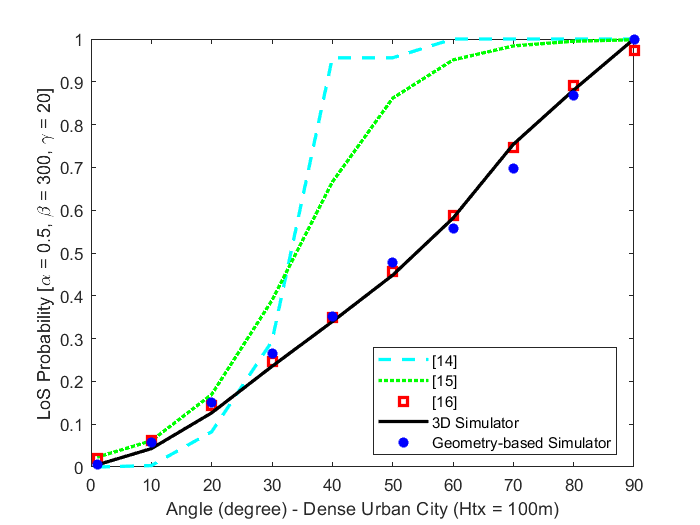}
    \caption{$P_{LoS}$ vs. $\theta$ (dense-urban).}
    \label{fig62}
  \end{subfigure}
  \centering
  \begin{subfigure}[b]{0.3\linewidth}
    \includegraphics[width=\linewidth]{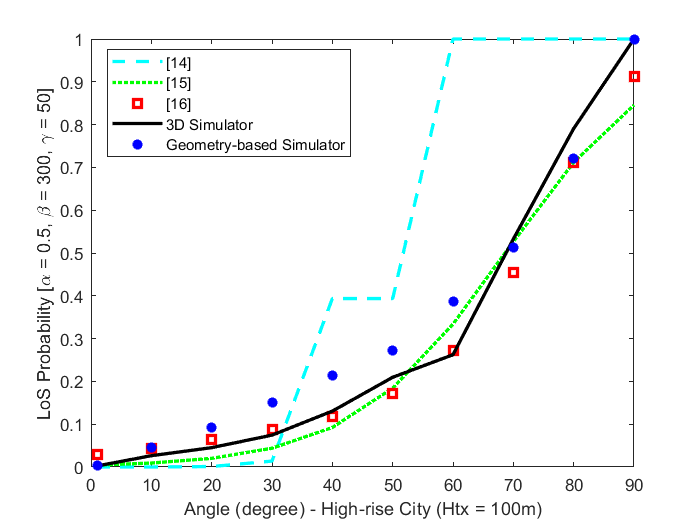}
    \caption{$P_{LoS}$ vs. $\theta$ (high-rise urban). }
    \label{fig63}
  \end{subfigure}
  \caption{$P_{LoS}$ as a function of elevation angle $\theta$ in different environments where \cite{ITU} is the ITU $P_{LoS}$ model that is step-wised with $\theta$, \cite{id13} only considers 2D scenario, i.e., a single row of buildings, and \cite{ICC} is limited to typical four urban environments.}
  
  \label{fig6}    
\end{figure*}

Both simulators can be used for reproducing any city environment using built-up parameters. The following Section discusses the results obtained with both simulators.

\begin{figure}[!t]
   \begin{minipage}{0.24\textwidth}
     \centering
     {\includegraphics[width=.95\linewidth]{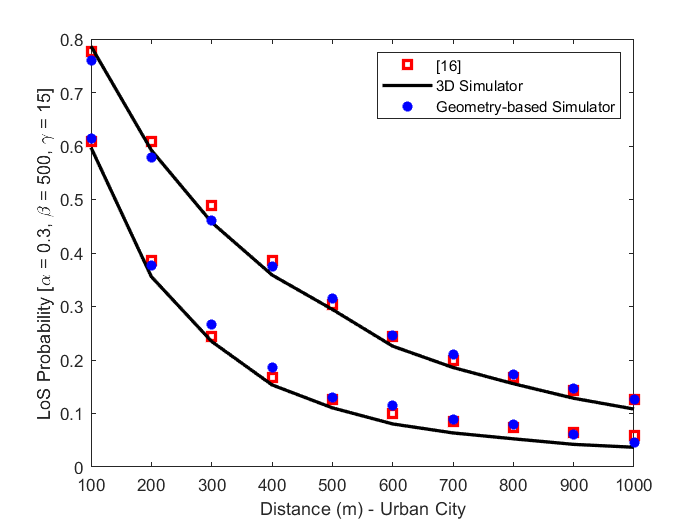}}
     \caption{$P_{LoS}$ as a function of radius in urban environment.}
     \label{fig7}
   \end{minipage}\hfill
   \begin{minipage}{0.24\textwidth}
     \centering
     {\includegraphics[width=0.95\linewidth]{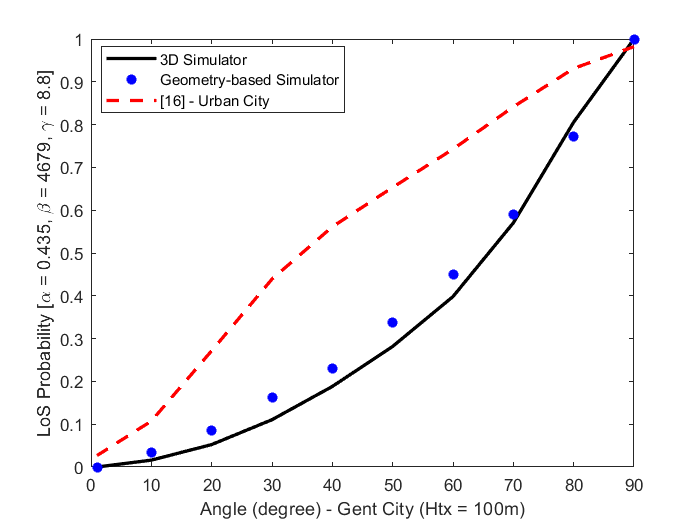}}
     \caption{Practical $P_{LoS}$ evaluation in Ghent (Belgium).}
     \label{fig8}
   \end{minipage}
\end{figure}


\section{Results and Analysis}
This section compares the $P_{LoS}$ of the proposed simulators with SOTA models \cite{ITU, id13, ICC}.
\paragraph{Accuracy for standard environments}
 Fig. \ref{fig6} plots the $P_{LoS}$ as a function of $\theta$ (in degrees) for typical urban environments using the above-mentioned techniques. For example, Fig. \ref{fig61} illustrates $P_{LoS}$ as a function of $\theta$ in an urban environment. For this analysis, the 3D city simulator creates a new city for each simulation using built-up parameters and randomly places a UAV at an arbitrary height of 100m. After that, the $P_{LoS}$ is calculated for 360 user locations (distributed in a circle with the same horizontal distance from the UAV) for a $\theta$. The simulations are repeated 500 times, and the results are averaged. 

In contrast, the geometry-based simulator does not create an entire city. It places a user randomly in the street/crossroad and a UAV at 100m with an arbitrary $\varphi$ for certain $\theta$. After, that it only creates buildings between the user and UAV. Lastly, it estimates $P_{LoS}$ averaged over 1000 runs. The results indicate a direct relation between $P_{LoS}$ and $\theta$ for simulators and SOTA models. The lower elevation angles between UAVs and users show less $P_{LoS}$. The primary reason for this trend is the larger distance between the UAV and the user at lower elevation angles, creating more obstructing buildings and shadowing. 

The proposed simulators' results are well aligned with each other, and \cite{ICC} (in Fig. \ref{fig61}, Fig. \ref{fig62} and Fig. \ref{fig63}). However, a significant mismatch exists with \cite{ITU,id13}. It is mainly due to the limitation of the proposed models in 2D (x and z-axis), as explained in Fig. \ref{fig1}. The same trends are followed in Fig. \ref{fig62} and Fig. \ref{fig63}. There is one significant mismatch in simulators' results, and \cite{ICC} in the high-rise scenario at $\theta = 90$. However, it does not make sense to have 91 \% $P_{LoS}$ when a UAV is flying straight above the user, as indicated by \cite{ICC}. 

Fig. \ref{fig7} plots the $P_{LoS}$ against the distance of the user. This figure also indicates that the results of both simulators are aligned and similar to \cite{ICC}, even when the UAV is flying at different altitudes. 

\textbf{Takeaway 1:} Popular State-of-the-art LoS models do not accurately represent their reference geometry (i.e., Manhattan grid city) and tend to significantly overestimate $P_{LoS}$. This can result in overestimating the potential of UAV-enabled applications. Only \cite{ICC} provides relatively accurate results, however, it is not widely used.     

\begin{figure}[!t]
     \centering
     {\includegraphics[width=.85\linewidth]{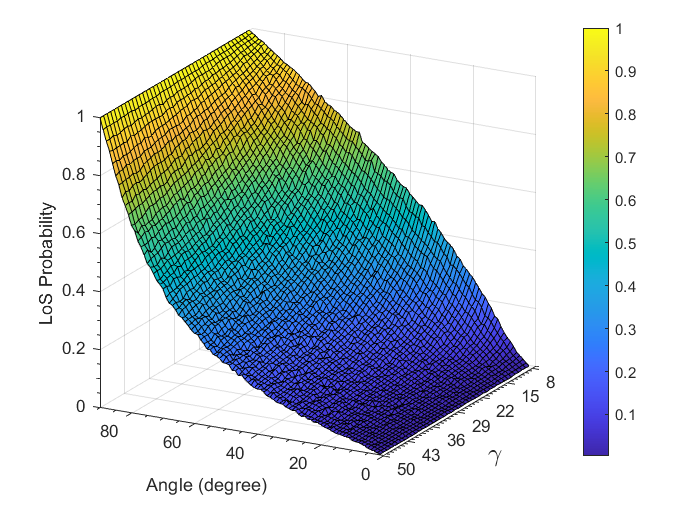}}
     \caption{$P_{LoS}$ as a function of $\gamma$ and $\theta$.}
     \label{fig9}
\end{figure}

\begin{figure}
     \centering
     {\includegraphics[width=.85\linewidth]{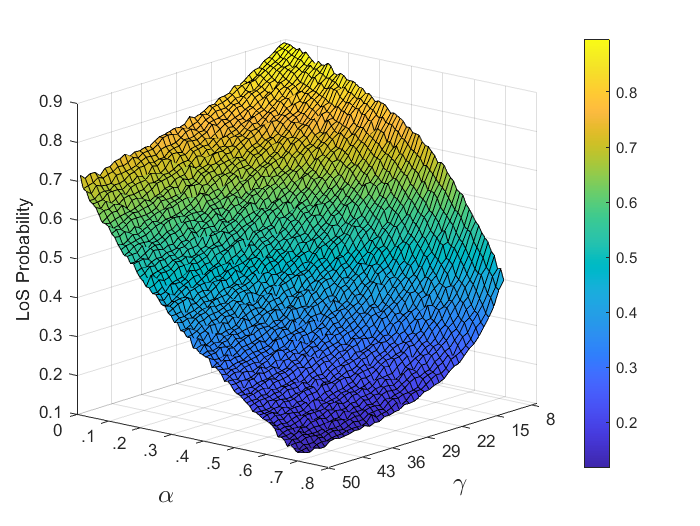}}
     \caption{$P_{LoS}$ as a function of $\gamma$ and $\theta$.}
     \label{fig10}
\end{figure}

\paragraph{Adaptability for arbitrary environments}
One major limitation of \cite{ICC} is that the model provides only four fitting parameters; therefore, it is limited to four typical urban environments mentioned in Table \ref{tab1}. The same limitation is observed in \cite{id13}. In contrast to SOTA models, the proposed simulators can work for any arbitrary environment, given built-up parameters. For analysis, we extract built-up parameters for Ghent (Belgium) city centre ($\alpha = 0.435, \beta = 4679, \gamma = 8.8$) and evaluate $P_{LoS}$. Fig. \ref{fig8} compares the obtained results with the most similar environment from \cite{ICC}. As one can observe, the difference is significant. Lastly, to demonstrate the viability of the proposed simulators within diverse environments using built-up parameters, we plotted $P_{LoS}$ as a function of ($\gamma$, $\theta$) and ($\alpha$, $\gamma$) in Fig. \ref{fig9} and Fig. \ref{fig10}, respectively. It is evident from both figures that the $P_{LoS}$ can easily be extracted for any tuple of ($\alpha, \beta, \gamma$). 


\textbf{Takeaway 2:} Existing LoS models have limited adaptability to changing build-up parameters as they rely on predefined sets of parameters 
 estimated empirically. 

\begin{figure}[!t]
\centerline{\includegraphics[width=.85\linewidth]{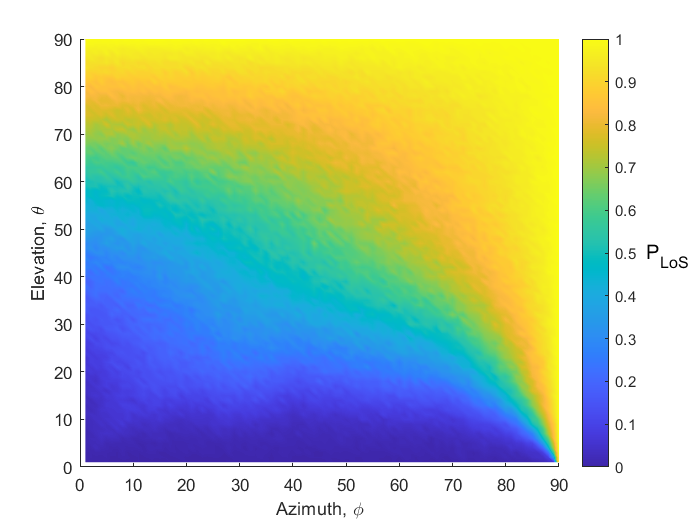}}
\caption{$P_{LoS}$ as a function of $\theta$ and $\varphi$ (UEs on the Street).  }
\label{3dplot}
\end{figure}

\begin{figure}[!t]
\centerline{\includegraphics[width=.85\linewidth]{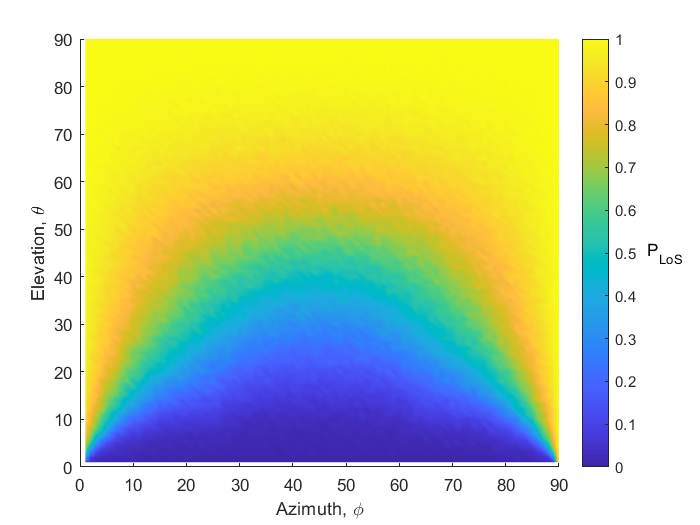}}
\caption{$P_{LoS}$ as a function of $\theta$ and $\varphi$ (UEs on the Crossroad).  }
\label{3dplotCR}
\end{figure}

\paragraph{Adaptability for different UE and UAV positions}
Locations of the nodes have a significant effect on $P_{LoS}$. Fig. \ref{3dplot} shows the behavior of $P_{LoS}(\theta, \varphi)$ for all the UEs on the street. We can observe significant variations over the azimuth. Indeed, when the user is located at Street 1 (see Fig.~\ref{fig5}) and $\varphi=0^{\circ}$, the scenario is well approximated by the geometrical representation depicted by Fig.~\ref{fig2} with $w$ and $s$ calculated as in \eqref{eq2} and \eqref{eq3}, respectively. However, when the azimuth is changed, the link geometry also changes. For example, the distance between two buildings is not equal to $s$ anymore (see Fig.~\ref{fig5}), which, in its turn, changes $P_{LoS}$. Note that when $\varphi=90^{\circ}$, $P_{LoS}$ becomes equal 1 since there are no buildings that can obstruct LoS. Alternatively, Fig. \ref{3dplotCR} illustrates the behavior of $P_{LoS}(\theta, \varphi)$ for all the UEs on the crossroad. In this scenario, $P_{LoS}$ equals 1 for $\varphi = 0^{\circ}$ 
and $90^{\circ}$ because of no obstructing buildings at these azimuth angles, as visualized in Fig.~\ref{fig5}. Hence, the location of UE or UAV and azimuth significantly impacts $P_{LoS}$ and must be considered for accurate estimation of $P_{LoS}$.   


\textbf{Takeaway 3:} While most of the SOTA works focus on the influence of the elevation angle, the effect of azimuth is not captured in any of the existing $P_{LoS}$ models. Therefore, the SOTA models cannot accurately predict the LoS behavior for different UAV or user locations with variable street lengths. 

\section{Conclusions}
This paper proposed a 3D city simulator for $P_{LoS}$ evaluation in an arbitrary city environment using built-up parameters. Furthermore, we presented a scalable and lightweight simulator to validate the 3D simulator results. Our analysis shows that existing models are less realistic and overestimate $P_{LoS}$, resulting in overestimating the potential of UAV-enabled applications. At the same time, such models work for a particular set of environments, which limits ABS analysis in any set of environments. In contrast, our simulators can be generalized for an arbitrary urban layout environment. Ultimately, we investigated the effect of azimuth on $P_{LoS}$, which is largely missing in SOTA models. Our findings show that $P_{LoS}$ varies with azimuth for a particular $\theta$, which calls for an urgent need to develop more realistic 3D $P_{LoS}$ models by incorporating azimuth to explore NTNs true potential.


\section*{Acknowledgements}
This research is supported by the Research Foundation Flanders (FWO), project no. G098020N, and KU Leuven Postdoctoral Mandate (PDM) under project no. 3E220691.
\bibliographystyle{IEEEtran}
\bibliography{ref}

\end{document}